\documentclass[11pt, a4paper]{article}

\usepackage[margin=1in]{geometry}

\usepackage{graphicx}
\usepackage{booktabs}
\usepackage{multirow}

\usepackage{amsmath,amssymb,amsfonts,amsthm}
\usepackage{mathrsfs}
\usepackage{textcomp}

\usepackage[T1]{fontenc}
\usepackage[utf8]{inputenc}
\usepackage{authblk}

\usepackage[colorlinks=true,linkcolor=blue,citecolor=blue,urlcolor=blue]{hyperref}

\usepackage{algorithm}
\usepackage{algorithmicx}
\usepackage{algpseudocode}
\usepackage{listings}

\theoremstyle{plain}

\theoremstyle{definition}

\begin{document}

\title{Evaluating Mechanical Property Prediction across Material Classes using Molecular Dynamics Simulations with Universal Machine-Learned Interatomic Potentials}

\author[1]{Konstantin Stracke}
\author[1]{Connor W. Edwards}
\author[1]{Jack D. Evans\thanks{Corresponding author: j.evans@adelaide.edu.au}}

\affil[1]{School of Physics, Chemistry and Earth Sciences, The University of Adelaide, North Terrace, Adelaide, 5005, South Australia, Australia}

\date{\today}

\maketitle

\begin{abstract}
We assess the accuracy of six universal machine-learned interatomic potentials (MLIPs) for predicting the temperature and pressure response of materials by molecular dynamics simulations. Accuracy is evaluated across 13 diverse materials (nine metal-organic frameworks and four inorganic compounds), computing bulk modulus, thermal expansion, and thermal decomposition. These MLIPs employ three different architectures (graph neural networks, graph network simulators, and graph transformers) with varying training datasets. We observe qualitative accuracy across these predictions but systematic underestimation of bulk modulus and overestimation of thermal expansion across all models, consistent with potential energy surface softening. From all tested models, three top performers arise; `MACE-MP-0a', `fairchem\_OMAT', and `Orb-v3', with average error across metrics and materials of $41~\%$, $44~\%$, and $47~\%$, respectively. Despite strong overall performance, questions arise about the limits of model transferability: dataset homogeneity and structural representation dominate model accuracy. Our results show that certain architectures can compensate for biases, a step closer to truly universal MLIPs.
\end{abstract}

\vspace{1em}

\section{Introduction}\label{Introduction}

The fundamental physics governing interatomic forces are well established, yet their application to complex systems yields equations that render quantum-mechanical methods such as density functional theory (DFT) computationally prohibitive\cite{dirac1929quantum}.
Machine-learning offers a promising alternative by learning the underlying potential energy surface (PES) from highly accurate reference data (often DFT), enabling near-ab initio accuracy at dramatically reduced cost\cite{behler2007generalized}.
The field has progressed toward an ambitious goal: the development of universal, general-purpose models capable of accurately capturing the essential physics of complex systems, without requiring system-specific parameterization or fine-tuning\cite{batatia2023foundation}.
This vision mirrors earlier efforts in classical force field development, from systematic system-specific parametrization approaches for coarse-grained models\cite{nguyen2022systematic} or atomistic force fields like QuickFF\cite{vanduyfhuys2018extension} and MOF-FF\cite{bureekaew2013mof}, toward broadly transferable frameworks exemplified by the Universal Force Field\cite{rappe1992uff}.
While recent studies have demonstrated that system specific fine-tuning of machine-learned interatomic potentials (MLIPs) can enhance accuracy for particular applications, for instance in high-temperature dynamics\cite{edwards2025exploring}, there is promise that universal MLIPs can offer robust `out-of-the-box' performance across a wide range of chemical environments and conditions.

Over time, graph-based MLIP architectures have become dominant\cite{batzner20223}, including graph neural networks (GNN) that estimate the atom (node) environment by passing messages from neighboring atoms within a cutoff\cite{unke2019physnet}.
Graph network simulators (GNS) extend this approach by incorporating dynamic properties such as velocities into the node embeddings\cite{sanchez2020learning}.
Graph transformers (GTs) further enhance message passing by weighting (`attention') contributions so that information can propagate beyond the local cutoff and potentially capture global effects\cite{musaelian2023learning}.

Mechanical and thermal properties are critical descriptors that quantify how materials respond to fundamental environmental stimuli.
Specifically, the bulk modulus quantifies resistance to compression, thermal expansion coefficients describe dimensional response to temperature changes, and decomposition temperatures define the stability to thermal stress.
Together, these properties define both the scope of practical applications and the operational limits of a material.
The computational speedup of MLIPs relative to DFT has enabled high-throughput materials screening\cite{batatia2023foundation}.
However, another transformative potential of these methods lies in the production of long-timescale molecular dynamics (MD) simulations.
This capability comes with a trade-off that MLIPs are physically agnostic\cite{morrow2023validate}, making benchmarking and validation a necessity.
Recently, Kraß et al. provided a systematic benchmark of mechanical properties, comparing MLIPs against DFT and classical potentials for metal-organic frameworks (MOFs)\cite{krass2025mofsimbench}.
They demonstrated that static property validation is insufficient and dynamic modeling must also be assessed, which was investigated by $NPT$ MD, including coordination environment and cell parameter analysis, focusing on MD stability.

In this work, we have evaluated the ability of universal MLIPs to predict key mechanical properties across a variety of material classes.
Six MLIP models were selected that span three graph-based architectures (GNS, GNN, GT) and trained on different datasets to investigate the interplay between model architecture, training data, and predictive performance.
This selection allows systematic trends to be assessed and the effects of architecture versus data coverage to be isolated.
For this study, a database comprising 13 materials with diverse and structurally complex characteristics was constructed (Table~\ref{tab:Intro_materials}).
The comprehensive scope of this study encompasses 13 materials with densities ranging from $0.19$ to $2.93$ g~cm$^{-3}$ across multiple material classes and compositions, demonstrating the breadth and complexity of the systems investigated.
A series of $NPT$ and $NVT$ MD simulations were used to evaluate the temperature- and pressure-dependent response of each material, including negative thermal expansion, bulk modulus, and thermal decomposition.
These simulation-derived properties were subsequently validated against reference data, prioritizing experimental measurements and supplemented with quantum mechanical calculations.

\begin{table*}[h!]
\centering
\caption{Materials investigated in this study, including their chemical formula, density, and classification, with MOFs further subdivided by structural or compositional family.}
    \label{tab:Intro_materials}
\begin{tabular}{l l c l}
\hline
Material & Chemical Formula & Density / g~cm$^{-3}$ & Class \\
\hline
\multicolumn{4}{l}{\textbf{Zn-based MOFs (carboxylate linkers)}} \\
MOF-5          & C$_{24}$H$_{12}$O$_{13}$Zn$_{4}$ & 0.35 & Zn-BDC (terephthalate) \\
MOF-10         & C$_{42}$H$_{24}$O$_{13}$Zn$_{4}$ & 0.19 & Zn-BPDC (biphenyl dicarboxylate) \\
TRUMOF       & C$_{24}$H$_{12}$O$_{13}$Zn$_{4}$ & 0.84 & Zn-truxene-BDC \\[3pt]
\multicolumn{4}{l}{\textbf{Zn-based MOFs (azolate linkers)}} \\
ZIF-8          & C$_{6}$H$_{6}$N$_4$Zn & 0.49 & ZIF (imidazolate) \\
CALF-20        & C$_{3}$H$_{2}$N$_3$O$_2$Zn & 0.94 & Zn-triazolate-oxalate \\[3pt]
\multicolumn{4}{l}{\textbf{Zr-based MOFs (UiO/NU family)}} \\
UiO-66         & C$_{24}$H$_{14}$O$_{16}$Zr$_3$ & 0.72 & Zr-BDC (terephthalate) \\
UiO-66-NH$_2$  & C$_{24}$H$_{17}$N$_3$O$_{16}$Zr$_3$ & 0.76 & Zr-NH$_2$-BDC \\
UiO-67         & C$_{42}$H$_{26}$O$_{16}$Zr$_3$ & 0.43 & Zr-BPDC (biphenyl dicarboxylate) \\
NU-1000        & C$_{44}$H$_{30}$O$_{16}$Zr$_3$ & 0.29 & Zr-TBAPy (pyrene tetrabenzoate) \\[3pt]
\multicolumn{4}{l}{\textbf{Inorganic materials}} \\
Zn(CN)$_2$     & C$_2$N$_2$Zn & 1.16 & Coordination polymer \\
SiO$_2$        & SiO$_2$ & 1.30 & Inorganic oxide, $\alpha$/$\beta$-Cristobalite  \\
Zr(WO$_4$)$_2$ & W$_2$ZrO$_8$ & 2.92 & Inorganic oxide \\
CaMn$_7$O$_{12}$ & CaMn$_7$O$_{12}$ & 2.87 & Perovskite oxide \\
\hline
\end{tabular}
\end{table*}

\section{Methods}\label{Method}
All $NVT$ and $NPT$ MD simulations were performed within the Atomic Simulation Environment\cite{larsen2017atomic} with six different MLIP models (Table~\ref{tab:model_specs}).
The models include the `Orb-v3' MLIP (Orb-v3-conservative-inf-omat)\cite{rhodes2025orb} from the Orb framework\cite{neumann2024orb}, which is trained on the OMat dataset\cite{barroso-luqueOpenMaterials2024}.
Multiple versions of the MACE framework (version 0.3.13) were tested, including `MACE-1' (MACE-MP-0a)\cite{batatia2023foundation}, trained on the Materials Project trajectories dataset\cite{MPtrj_2023}, and `MACE-2’ (MACE-MPA-0), an updated version trained on an expanded dataset that includes the subsampled Alexandria dataset\cite{salex_2024}.
Additionally, the `MACE-MOF' was included, which is the  MACE-MP-0b (medium) model fine-tuned for metal-organic frameworks (MOF) and fine-tuned with a dataset of 127 MOFs\cite{elena2025machine}.
Finally, from the Fairchem framework\cite{fairchem2025}, two pre-trained models were tested;
the 'OMat' model trained on the OMat dataset\cite{barroso-luqueOpenMaterials2024} and the 'ODAC' model specifically trained for MOFs and direct air capture applications\cite{sriram2025odac25}.

\begin{table*}[h!]
    \centering
        \caption{Summary of the MLIP models employed with their respective DOI. Further the MLIPs class of Graph Network Simulator (GNS), Graph Neural Network (GNN) or Graph Transformer (GT) and the training datasets are also shown.}
    \label{tab:model_specs}
    \begin{tabular}{lccl}
        \toprule
        Name & DOI &Class& Datasets\\
        \midrule
        Orb-v3       & 10.48550/arXiv.2504.06231 & GNS & OMat\cite{barroso-luqueOpenMaterials2024}\\
        MACE-1        & 10.48550/arXiv.2401.00096 &GNN & MPtrj\cite{MPtrj_2023}\\
        MACE-2        & 10.48550/arXiv.2401.00096  &GNN& MPtrj\cite{MPtrj_2023} + sAlex\cite{salex_2024}\\
        MACE-MOF      & 10.1038/s41524-025-01611-8 & GNN& MPtrj\cite{MPtrj_2023} + 127 MOFs\\
        fairchem\_OMAT & 10.5281/zenodo.15587498 & GT&OMat\cite{barroso-luqueOpenMaterials2024}\\
        fairchem\_ODAC & 10.5281/zenodo.15587498 & GT &ODAC\cite{sriram2025odac25}\\
        \bottomrule
    \end{tabular}
\end{table*}

$NVT$ simulations at ambient temperature used the cell and geometry optimized configuration which was isotropically scaled to volumes corresponding to $0.96, 0.98, 1.00, 1.02$, and $1.04$ times the original volume of the cell and subsequently geometry optimized.
$NPT$ simulations were performed at temperatures ranging from $50$ to $1000~\text{K}$ in $50~\text{K}$ increments.
Each temperature step was initialized from the final configuration of the preceding step, with the $50~\text{K}$ simulation starting from the optimized structure.
Both $NVT$ and $NPT$ simulations ran for a total of $125~\text{ps}$ with a timestep of $1~\text{fs}$, and system configurations were recorded every 50 steps, which were used for all subsequent analyses.
The initial $25~\text{ps}$ of the simulation trajectory was reserved for equilibration.
Temperature and pressure were controlled using the Berendsen thermostat and barostat\cite{berendsen1984molecular}.
The thermostat used a  timestep of $1~\text{fs}$ and a temperature coupling ($\tau_T$) of $100~\text{fs}$, while the barostat time constant ($\tau_p$) and the material compressibility were adjusted for each simulation, as described in the Supplementary Information (Section~S3.2).

The bulk modulus at constant temperature $K_T$ is defined as the volume $V$ that changes with a pressure change $\partial P$:
\begin{align}
K_T = -V\left(\frac{\partial P}{\partial V}\right)_T
\label{eq1}
\end{align}
In practice, $K_T$ is calculated from $NPT$ simulations via averaged pressures from the stress tensor components.
Bulk modulus was then determined using linear regression over the pressure–volume data collected at different temperatures.
In contrast to $NPT$, the $NVT$ simulations of bulk moduli were calculated by the Birch-Murnaghan pressure-volume relationship fit.
For calculating the bulk modulus values, only $NVT$ and $NPT$ simulations at $300~\mathrm{K}$ were used to compare to literature values also measured at $300~\mathrm{K}$.
To show these simulations are equilibrated, autocorrelation functions for volume ($NPT$) and pressure ($NVT$) at $300~\mathrm{K}$ are provided in the Supplementary Information (Sections S2.1 and S3.1).

The volumetric coefficient of thermal expansion (CTE) $\alpha_V$, is defined as the change in volume with respect to temperature at constant pressure.
\begin{equation}
\alpha_V = \frac{1}{V}\left(\frac{\partial V}{\partial T}\right)_P 
\label{eq3}
\end{equation}
For each material and model, $\alpha_V$ was obtained from the slope of the volume–temperature relationship, computed from the $NPT$ simulations by evaluating the temperature dependence of the average cell volume and averaged over the reference temperature range.

The temperature of decomposition ($T_{\mathrm{decomp.}}$) was defined by coordination bond breakage between the metal-ligand bond.
First, organic atoms bonded to the metals are identified from the optimized structures using a predefined cutoff corresponding to the equilibrated bond length.
An additional threshold of $1~\text{Å}$ was added to the equilibrium bond length to accommodate typical MD fluctuations at elevated temperatures.
The entire trajectory was then scanned and if any bond exceeds this threshold for $250~\text{fs}$, the material is classified as collapsed at that temperature.
Reference values for bulk modulus, thermal expansion, and thermal decomposition were compiled from experimental measurements and, where experimental values were unavailable, quantum mechanical calculations (Table~\ref{tab:references}).
These reference values are used to validate and evaluate the model accuracy of property predictions.

\begin{table*}[h!]
    \centering
        \caption{Reference values for all materials for the bulk modulus, ($K_T$), thermal expansion coefficient ($\alpha_V$), with temperature range, and the thermal decomposition temperature.}
    \begin{tabular}{l|c|cc|c}
        \toprule
        Material & Bulk Modulus &  \multicolumn{2}{c|}{Thermal Expansion}     & Decomp. \\
         Name & $K_T$ / GPa                  &$\alpha_V$ / MK$^{-1}$& $\Delta T$ / K & \multicolumn{1}{c}{$T_{\mathrm{decomp.}}$ / K}\\            
         \midrule
         MOF-10          &  6.0  \cite{wieser2024machine}    & -65  \cite{dubbeldam2007exceptional}  & 80-500    &   650   \cite{chung2017influence}    \\
         MOF-5           &  17.0 \cite{mattesini2006ab}     & -48  \cite{zhou2008origin}  & 4-600     &        673-783 \cite{healy2020thermal} \\
         ZIF-8           & 6.5  \cite{chapman2009pressure}     & 27.6 \cite{chester2024loading} & 200-500   & 550     \cite{park2006exceptional}    \\
         NU-1000         &  8.2  \cite{robertson2023survival}    & -42 \cite{chen2022node}   & 430-450   &    700    \cite{mondloch2013vapor}  \\
         CALF-20         & 29.1   \cite{fan2024unconventional}  & 11.54 \cite{fan2024unconventional}  & 10-400   &   380    \cite{phongsuk2025efficient}   \\
         TRUMOF          & 7.5 \cite{meekel2024enhanced}      & -      & -        &   450    \cite{meekel2023truchet}   \\
         UiO-66          &  41.0  \cite{wu2013exceptional}   & -35 \cite{vornholt2024node}   & 50-250   &     650  \cite{valenzano2011disclosing}   \\
         UiO-66-NH$_2$   & 17.0  \cite{burtch2018mechanical}    & -       & -        &  400  \cite{garibay2010isoreticular} \\
         UiO-67          &  17.4  \cite{hobday2016computational}  & -26.1 \cite{goodenough2020interplay} & 550-800    & 670 \cite{goodenough2020interplay}   \\
         Zn(CN)$_2$   &  34.2\cite{chapman2007pressure} &-51\cite{goodwin2005negative} &10-370 &375\cite{goodwin2005negative} \\
         Zr(WO$_4$)$_2$  &  61.3   \cite{perottoni1998pressure}  & -26.7 \cite{mary1996negative} & 0.3-1050  &   1050\cite{mary1996negative}   \\
         CaMn$_7$O$_{12}$& 190   \cite{stekiel2022pressure}    & 19.5  \cite{gautam2017large} & 377-488   & \>550   \cite{gautam2017large} \\
         SiO$_2$ ($\beta$-cristobalite)         &      37\cite{ricci2017ab}   & 32.7\cite{aumento1966stability} & 100-500  & 1986 \cite{haynes2016crc}   \\
         \bottomrule
    \end{tabular}

    \label{tab:references}
\end{table*}

The densities in Table~\ref{tab:Intro_materials} are obtained at $300$~K, averaging over all models and their corresponding MD trajectories.
Because CTE values depend strongly on temperature, comparisons between reference data and simulations are performed only within the temperature range specified by the reference source.
When comparing models using the mean absolute error (MAE, \%), CaMn$_7$O$_{12}$ and Zr(WO$_4$)$_2$ were omitted because Ca and W are not represented within MACE-MOF.

\section{Results and Discussion}\label{Results}
\begin{figure*}[h!]
\centering
\includegraphics[width=1.0\textwidth]{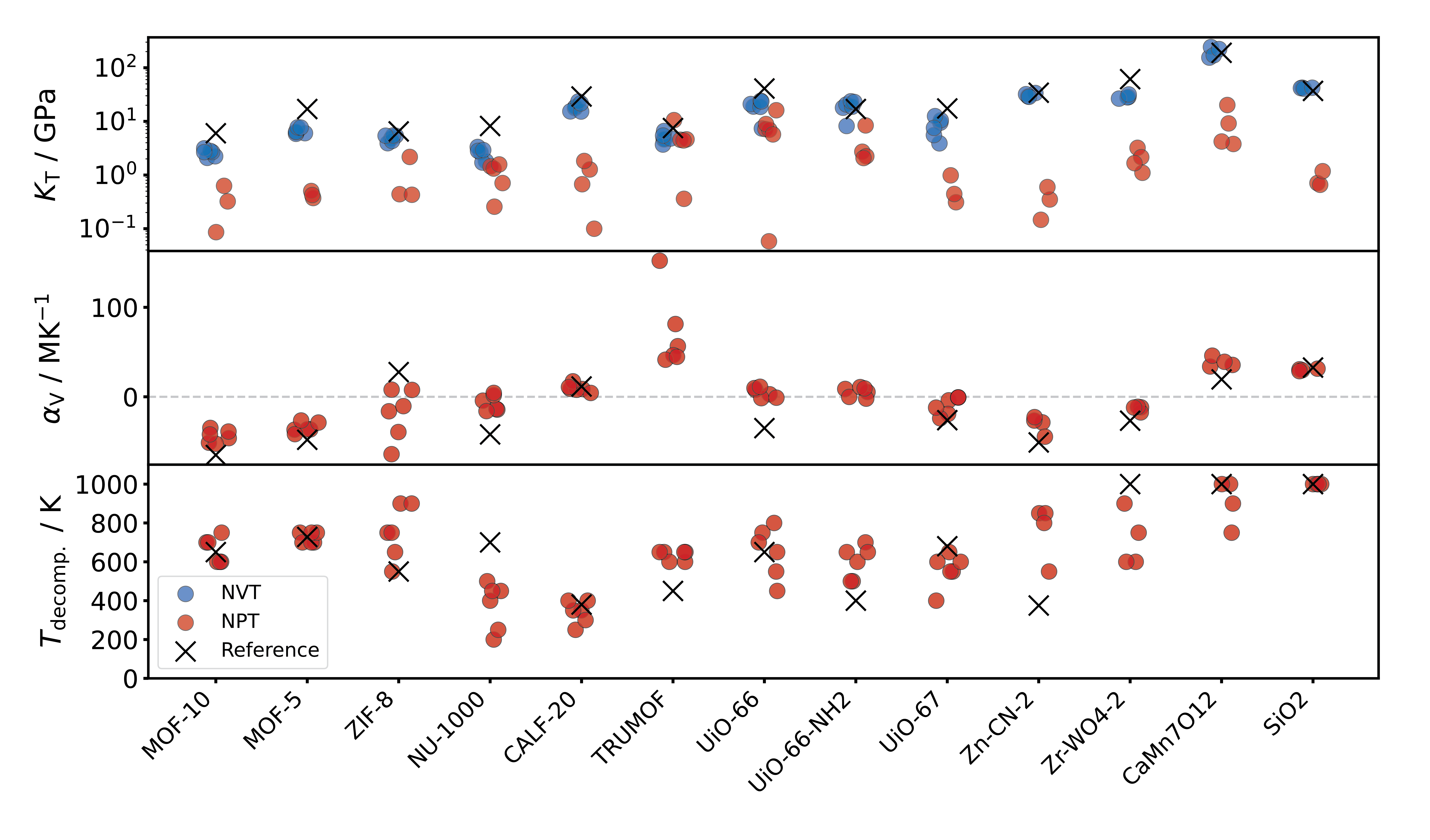}
\caption{Summary of the bulk modulus ($K_T$), volumetric thermal expansion coefficient ($\alpha_V$), and decomposition temperature for all materials, as predicted by the models. Corresponding literature values (Table~\ref{tab:references}) are highlighted for direct comparison.}
\label{fig:Bulk}
\end{figure*}

The behavior of materials, from microscopic atomic rearrangements to macroscopic deformations, is fundamentally shaped by temperature and pressure, the most basic external stimuli.
Bulk modulus, the material response to pressure, has macroscopic implications for engineering, among other fields, and it directly arises from the atomistic structure.
With its strong dependence on the atomistic structure and the physical and chemical environment, the bulk modulus is a central property in materials science.
In Figure~\ref{fig:Bulk} the calculated bulk moduli $K_T$ from two different approaches, $NVT$ and $NPT$ simulations, with respective reference values are presented.

Overall, the bulk modulus from $NVT$ simulations matches the qualitative trends between materials while also achieving an MAE of $43.8~\% \pm 6.9~\%$, relative to reference values.
As outlined in the Supplementary Information (Sections S2.1 and S3.1) the autocorrelation analysis of the MD trajectories show that volume relaxes substantially more slowly than pressure.
Averaged over all materials at 300 K, the characteristic correlation time of volume is $6.72~\text{ps}$, compared to $59.5~\text{fs}$ for pressure, indicating that volume relaxes approximately $114$ times slower.
Thus, the bulk modulus was more reliably obtained from short NVT trajectories\cite{wittmer2013compressibility}.
This is clearly illustrated by the example of CaMn$_7$O${_{12}}$, where the $NPT$ simulations yield an average value of $9.3~ \text{GPa}$, while the $NVT$ simulations give $197.8~ \text{GPa}$, which is much closer to the experimentally-determined value of $190~\text{GPa}$\cite{stekiel2022pressure}.

Many MOFs exhibit fascinating negative thermal expansion (NTE), which can be largely attributed to porosity and flexible linkers\cite{evans2019assessing}.
However, this behavior is not unique to MOFs and also occurs in Zr(WO$_4$)$_2$, which has rigid units allowing flexible rotational motion (`hinging') around these corners (Table~\ref{tab:references})\cite{miller2009negative, mary1996negative}.
Complicating matters, many materials undergo various forms of structural reordering, resulting in complex thermal behavior with different CTEs across temperature ranges, including transitions from positive to negative thermal expansion\cite{stracke2025investigating}.
Even when comparing within the same temperature range ambiguities remain.
For example, Vornholt et al. observed substantial thermal hysteresis in UiO-66, with CTE values ranging between $45$ and $-80~\text{MK}^{-1}$ depending whether the material is heated or cooled over the same temperature range.
In the same work, they determined the `intrinsic' equilibrated CTE to be at $-35~\text{MK}^{-1}$\cite{vornholt2024node}.
Additionally, universal MLIPs are not trained on higher temperature dynamics, which may improve their CTE predictions\cite{edwards2025exploring}.
As expected, predicting CTEs is more difficult compared to the bulk modulus ($43.8~\%$) with a MAE of $76.2~\% \pm 25.2~\%$.
Yet overall, the qualitative trend between materials and positive to negative thermal expansion is captured across models, with the exceptions of {UiO-66}, and for {ZIF-8} where only MACE-1 and {fairchem\_OMAT} correctly reproduce this behavior (Figure~\ref{fig:Bulk}).

Beyond a volumetric response to temperature, thermal decomposition temperature is a fundamental property, determining the suitability of a material for practical applications.
It reflects the ability to maintain structural integrity when subjected to heat, the resistance to bond breakage, and allows interatomic forces and bond-breaking to be analyzed using MLIPs.
Within similar types of materials, {MOF-10} is predicted to be less stable than {MOF-5}, while the UiO family is correctly predicted to be stable within a very similar temperature range.
Considering the intrinsic material complexity, this level of agreement is remarkable.
This in turn reinforces our approach of metal-organic bond breaking as an indicator for decomposition and reveals the ability of all tested MLIPs to accurately capture bond-breaking processes, in contrast to many classical force fields.

\begin{figure*}[h!]
\centering
\includegraphics[width=1\textwidth]{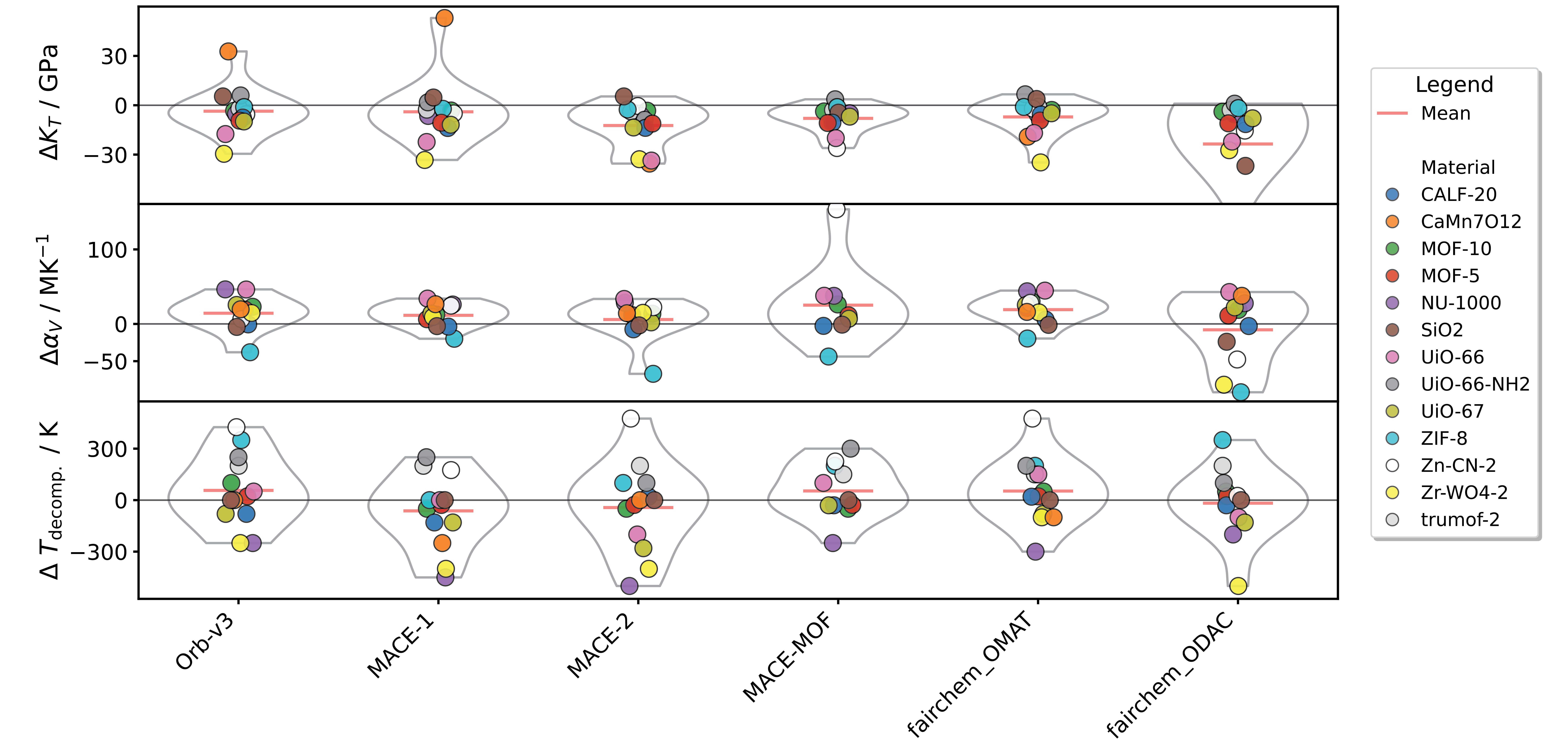}
\caption{Prediction deviations for bulk modulus $K_T$, CTE $\alpha_V$, and decomposition temperature across models, with materials indicated in the legend. Materials for which no reference value is available are omitted.}\label{fig:violin}
\end{figure*}

Beyond examining individual predictions, assessing the average accuracy across these predictions is essential and can highlight systematic errors.
In Figure~\ref{fig:violin} a violin plot is displayed summarizing the distribution of deviations for each metric from the reference.
The median predictions across all models are consistently close to the reference values, underscoring the overall accuracy of these models.
The median deviations from reference values, averaged across all models, are $-6.92~\text{GPa}$ for bulk modulus, $11.38~\text{MK$^{-1}$}$ for CTE and $18.50~\text{K}$ for decomposition temperature.
To put those absolute values in context, different metrics span different ranges, namely CTE $-64.10$ to $152.09~\text{MK}^{-1}$ ($\Delta~216.19~\text{MK}^{-1}$) and  for $K_T$ from $-11.21$ to $243.02~\text{GPa}$ ($\Delta~254.23$~\text{GPa}) and for decomposition temperature $200$ to $1000~\text{K}^{-1}$ ($\Delta~800~\text{K}^{-1}$).
This shows high accuracy when averaged across materials, but also reveals a systematic bias: all models underestimate the bulk modulus while overestimating the thermal expansion coefficient.
This behavior is consistent with a systematic softening of the learned PES\cite{deng2024overcoming}.
A softened PES reduces the predicted pressure, since pressure is the first derivative of energy with respect to volume, and at the same time permits larger volume fluctuations as thermal energy encounters smaller barriers.

A closer look of our simulations illustrates that specific materials challenge these models.
For bulk modulus, CaMn$_7$O${_{12}}$ emerges as an outlier and is overestimated by Orb-v3 and MACE-1 while underestimated by fairchem\_ODAC.
In this case, the material has an exceptionally large bulk modulus, which amplifies these absolute deviations.
Regarding CTE, {ZIF-8} is consistently the most underestimated material across all models.
For decomposition temperature, {NU-1000} and {Zr(WO$_4$)$_2$} appear to be the least accurately predicted materials.
While {NU-1000} is an anisotropic framework featuring a large tetradentate linker, structural changes in Zr(WO$_4$)$_2$ stem from the aforementioned `hinging' of rigid units.
These results demonstrate that universal MLIPs have reached a level of maturity where they can reliably predict complex thermomechanical properties across diverse material classes.
Although individual predictions can show deviations, the models exhibit exceptional qualitative accuracy when compared across material averages.
The systematic nature of observed biases, including an underestimation in bulk modulus and an overestimation in thermal expansion suggests clear pathways for future model refinement, possibly through targeted training data augmentation.

\begin{figure}[h!]
\centering
\includegraphics[width=0.5\textwidth]{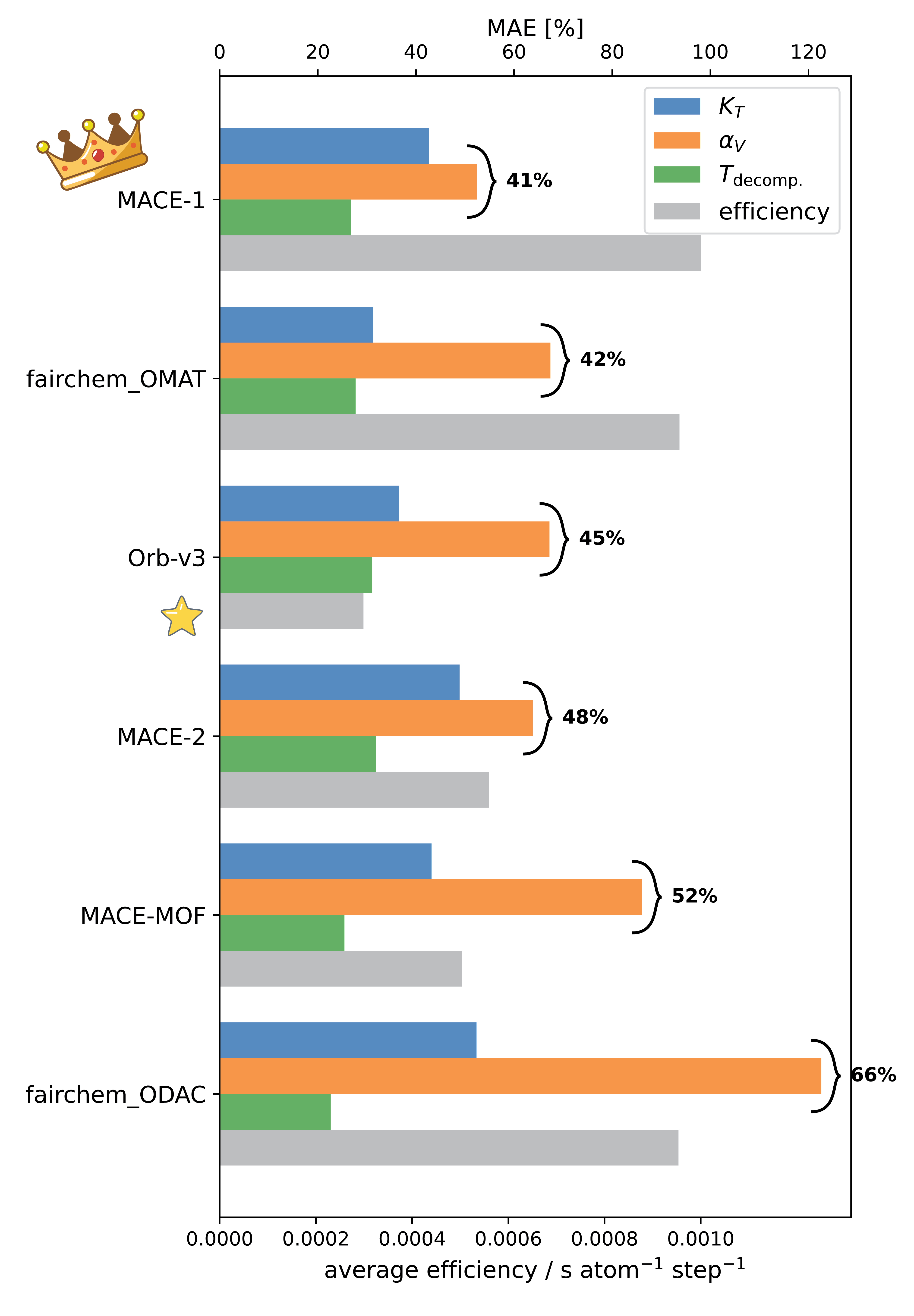}
\caption{Mean absolute error (MAE, $\%$) of all models across bulk modulus $K_T$, CTE $\alpha_V$, decomposition temperature, and efficiency. The average accuracy across metrics (excluding efficiency) is indicated.}\label{fig:score}
\end{figure}

Having established the predictive capabilities of MLIPs across the material set, the overall accuracy of the models was evaluated.
Figure~\ref{fig:score} compares the MAE ($\%$) of all models across the evaluated metrics in addition to efficiency and the average accuracy across all metrics.
The three best performing models are MACE-1, fairchem\_OMAT, and Orb-v3.
Fairchem\_OMAT has the highest accuracy for bulk, Orb-v3 is the fastest model, and MACE-1 has the highest accuracy across metrics.

Among these top three models (MACE-1, fairchem\_OMAT, and Orb-v3), the architectures are distinct, corresponding to a GNN, GT, and GNS, respectively.
Despite these architectural differences, all three achieve comparable overall accuracy.
The consistent accuracy of these three models across all materials, despite architectural diversity, underscores that data curation and coverage exert a greater influence on average performance than architectural approach.
We note this observation aligns with findings of Kraß et al.\cite{krass2025mofsimbench}.
Interestingly, two of the three best performers were trained on OMAT, while the top-performing model, MACE-1, was trained on MPtrj.
The OMAT-trained models excel at bulk modulus predictions but underperform on CTE.
This accuracy trade-off reflects the inherent characteristics of each training dataset.
OMAT contains non-equilibrium structures well-suited for molecular dynamics and bulk property predictions\cite{krass2025mofsimbench}, whereas MPtrj has relaxation trajectories which better capture vibrational modes.
Although fairchem\_ODAC performs poorly on vibrational properties and bulk modulus, its superior thermal stability predictions reflect training emphasis on accurate interatomic forces from MOF-guest structures.

We now examine the different MACE models: MACE-1, the `multi-head' fine-tuned MACE model\cite{elena2025machine}, MACE-MOF, and the MACE model that was trained on an enlarged dataset combining MPtrj and sAlex\cite{batatia2023foundation}, MACE-2.
Counterintuitively, the initial release MACE model (MACE-1) performs better than either.
This is surprising as the later model, MACE-2, was trained on more data and not fine-tuned.
Critically, both models share identical architectures\cite{batatia2023foundation}, eliminating architectural differences as an explanation.
Although MACE-2 has a lower accuracy than MACE-1, it achieves a higher computational efficiency, a trade-off that is also reported for MOF deformation by guests\cite{brabson2025comparing}.
The speed-up of MACE-2 stems from optimizations to how the architecture is implemented rather than architectural simplification.
Additionally, both training sets were generated by PBE/PBE+U functionals, which suggests it is not necessarily the level of theory that leads to less accurate results.
The sAlex structures (10.4M of 1D, 2D, and 3D inorganic materials\cite{schmidt2024improving}) vastly outnumber MPtrj structures (1.58M) in the combined training set.
The inclusion of additional structures from the sAlex dataset reduced predictive accuracy for structures already well represented in the MPtrj dataset.
Comparing MACE-MOF with MACE-1, the results show that they have similar accuracy when describing MOFs, however MACE-1 can also perform on non-MOF structures (Supplementary Figure~S1).
For MACE-MOF, the MPtrj workflow is used to generate DFT training data for the curated MOF dataset, with only D3 dispersion correction added, as this correction is deemed particularly important for MOFs.
While there are different approaches to fine-tuning\cite{allen2024learning}, the MACE-1 versus MACE-2 comparison demonstrates that mixing datasets can compromise performance.
In summary, underrepresented structural characteristics will lead to reduced model accuracy.
Homogeneous datasets, including both equilibrium and non-equilibrium structures sampled at a consistent level of theory, are needed for accurate dynamics and property prediction.

\begin{figure}[h!]
\centering
\includegraphics[width=0.5\textwidth]{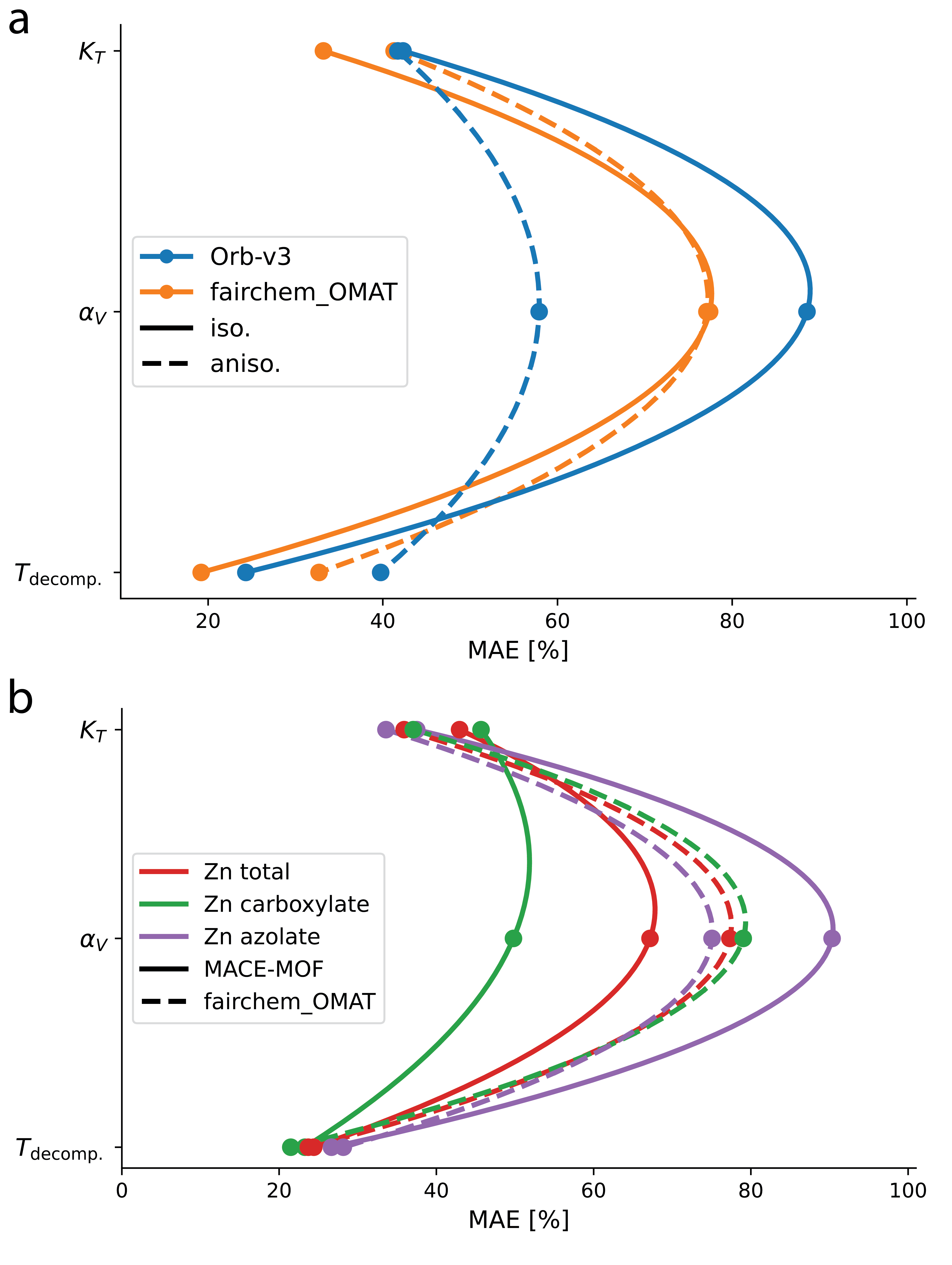}
\caption{Model performance for predicting bulk modulus ($K_T$), CTE ($\alpha_V$), and decomposition temperature across selected material subsets. a) Orb-v3 and fairchem\_OMAT evaluated on isotropic (iso.) and anisotropic (aniso.) MOFs. b) MACE-MOF and fairchem\_OMAT evaluated on zinc-based MOFs separated by linker type (Zn carboxylate vs. Zn azolate). Connecting curves added to aid visual interpretation.}
\label{fig:ISO_structure}
\end{figure}

To assess whether network architecture influences performance for specific systems rather than on average, we divided our MOF dataset based on cell symmetry into isotropic and anisotropic subsets ({NU-1000}, {CALF-20}, {UiO-66-NH$_2$}).
Subsets were curated according to the anisotropic thermal expansion behavior revealed by $NPT$ simulations (Supplementary Figure~S2).
Since fairchem\_OMAT and Orb-v3 were trained on the same dataset, performance differences can be attributed to their architectures or implementations.
In comparison, we find that fairchem\_OMAT performs slightly worse for anisotropic MOFs, a trend that holds across bulk modulus, CTE, and decomposition temperature (Figure~\ref{fig:ISO_structure}a).
Interestingly, Orb-v3 predicts the CTE of anisotropic MOFs more accurately than for isotropic MOFs.
This implies that Orb-v3 accuracy varies between isotropic and anisotropic MOFs, whereas fairchem\_OMAT shows consistent accuracy across both classes.
Collectively, all models except fairchem\_OMAT display this symmetry bias (Supplementary Figure~S3).

Finally, the influence of material chemistry on model performance was examined by focusing on zinc-based MOFs, separated into `Zn carboxylate' and `Zn azolate' subsets (Table~\ref{tab:Intro_materials}).
For this we compare the MOF fine-tuned model MACE-MOF against fairchem\_OMAT.
Figure~\ref{fig:ISO_structure}b shows that both models predict decomposition temperature with comparable accuracy across both linker types.
Both models predict bulk modulus slightly better for azolate frameworks, but the most pronounced difference emerges for CTE, where {MACE-MOF} achieves substantially higher accuracy for Zn–carboxylate MOFs.
The MACE-MOF model shows a strong bias toward Zn-carboxylate MOFs, achieving nearly twice the accuracy of Zn-azolate systems, with MAEs of $50~\%$ and $90~\%$, respectively.
This trend, with respect to CTE, is present for all models (Supplementary Fig.~S4), except fairchem\_OMAT, which shows near no changes between these ligands (Fig.~\ref{fig:ISO_structure}).
Notably, fairchem\_ODAC, with a different and more specific training dataset, displays the same bias.
The fairchem\_OMAT model achieves more uniform accuracy across ligand types and symmetries, despite using the same training data as {Orb-v3} and lacking the domain-specific fine-tuning of {MACE-MOF}, suggesting that thoughtful architectural design may be as important as training data or fine-tuning in mitigating prediction bias.

\section{Summary and Outlook}\label{Conclusion}
Our results reveal that universal MLIPs can reliably predict thermomechanical properties, though performance depends critically on both architecture and training data composition.
We have demonstrated detailed analysis across isotropic and anisotropic MOFs, as well as Zn carboxylate and azolate linkers, which reveal that GNN and GNS architectures exhibit biases toward these structural and symmetry differences, while GT appears unbiased.
By comparing fairchem\_OMAT to Orb-v3, which are both trained on the OMAT dataset, this bias can be isolated as architecture-dependent.
However, dataset quality supersedes architectural advantages: inconsistent data representation degrades predictive accuracy regardless of architecture (fairchem\_OMAT versus fairchem\_ODAC), while simultaneously reintroducing these structure and symmetry biases even in graph transformer models.
This principle extends to dataset modification strategies, enlarging the MPtrj dataset with sAlex data (MACE-2 vs. MACE-1) and fine-tuning on MOF-specific data (MACE-MOF vs. MACE-1) both failed to improve predictions.
While MACE-MOF showed no performance gains, even on its target MOF domain, both modified models exhibited degraded overall accuracy.

The top-performing models, MACE-1, fairchem\_OMAT, and Orb-v3, exhibit comparable overall accuracy despite distinct architectures (GNN, GT, and GNS, respectively), highlighting that dataset composition and coverage strongly influence performance.
Qualitative trends are well recreated across properties and the accuracy across materials and models is close to reference values, averaging at MAE of $44~\%$ for our top performers.
Yet, it is important to highlight that task specific trained models like fairchem\_ODAC have a low overall accuracy (MAE, $66~\%$).
However, they excel at task specific accuracy, namely predicting the decomposition temperature with a MAE of $23~\%$.
There is a systematic underestimation of bulk modulus and overestimation of thermal expansion across all models, consistent with PES softening.
Notably, all MD simulations remained stable, confirming the robustness of universal MLIPs even at elevated temperatures and deformed simulation cells.

From a practical standpoint, Orb-v3 demonstrates the fastest computational performance, while implementation optimizations in MACE-2 have achieved substantial speed improvements over its predecessor.
These advances in computational efficiency, combined with our findings on dataset quality and architectural bias mitigation, point beyond semi-universal, task-specific models toward the prospect of truly universal MLIPs.

\section*{Supplementary Information}
Plots showing accuracy of MACE-1 and MACE-MOF on MOF and non-MOF subsets, Model-averaged linear thermal expansion across all materials, model resolved accuracy for MOF subsets of Zn-Carboxylate and Zn-azolate, pressure and volume fluctuations ($NVT$ and $NPT$), pressure and volume autocorrelation functions ($NVT$ and $NPT$), barostat parameter refitting.
All code and data for this research is available on Zenodo: \href{https://doi.org/10.5281/zenodo.17730688}{10.5281/zenodo.17730688}.

\section*{Acknowledgements}
J.D.E is the recipient of an Australian Research Council Discovery Early Career Award (project number DE220100163) funded by the Australian Government.
Phoenix HPC service at the University of Adelaide is thanked for providing high-performance computing resources.
This research was supported by the Australian Government’s National Collaborative Research Infrastructure Strategy (NCRIS), with access to computational resources provided by Pawsey Supercomputing Research Centre through the National Computational Merit Allocation Scheme.

\end{document}